\documentclass[twocolumn,prd,amsmath,superscriptaddress]{revtex4}

\usepackage{hyperref}
\usepackage{ifthen}
\usepackage{amsmath}
\usepackage{graphics}
\usepackage{color}
\definecolor{darkgreen}{cmyk}{0.85,0.2,1.00,0.2}
\definecolor{purple}{cmyk}{0.5,1.0,0,0}

\def\simgt{\gtrsim}

\begin{document}

\title{Cosmological constraints on DGP braneworld gravity with brane tension}

\author{Lucas Lombriser}
\affiliation{Institute for Theoretical Physics, University of Z\"{u}rich, Winterthurerstrasse 190, CH-8057 Z\"{u}rich, Switzerland}
\author{Wayne Hu}
\affiliation{Kavli Institute for Cosmological Physics, Department of Astronomy and Astrophysics,Enrico Fermi Institute, University of Chicago, Chicago, Illinois 60637, USA}
\author{Wenjuan Fang}
\affiliation{Department of Physics, Columbia University, New York,
New York 10027, USA} \affiliation{Brookhaven National Laboratory, Upton, New York
11973, USA}
\author{Uro\v{s} Seljak}
\affiliation{Institute for Theoretical Physics, University of Z\"{u}rich, Winterthurerstrasse 190, CH-8057 Z\"{u}rich, Switzerland}
\affiliation{Physics and Astronomy Department, University of California, and Lawrence Berkeley National Laboratory, Berkeley, California 94720, USA}
\affiliation{Ewha University, Seoul 120-750, Korea}

\date{\today}

\begin{abstract}
We perform a Markov chain Monte Carlo analysis of the
self-accelerating and normal branch of Dvali-Gabadadze-Porrati
braneworld gravity.  By adopting a parametrized post-Friedmann
description of gravity, we utilize all of the cosmic microwave
background data, including the largest scales, and its correlation
with galaxies in addition to the geometrical constraints from
supernovae distances and the Hubble constant. We find that on both
branches brane tension or a cosmological constant is required at
high significance with no evidence for the unique
Dvali-Gabadadze-Porrati modifications. The crossover scale must
therefore be substantially greater than the Hubble scale $H_0 r_c >
3$ and 3.5 at the 95\% C.L. with and without uncertainties from
spatial curvature.  With spatial curvature, the limit from the
normal branch is substantially assisted by the galaxy cross correlation which highlights its importance in constraining infrared
modifications to gravity.
\end{abstract}

\maketitle

\section{Introduction}

Cosmological tests of the acceleration of the expansion offer unique opportunities to test
gravity on large scales and low curvature.
Dvali, Gabadadze, and Porrati (DGP)~\cite{dvali:00} proposed that such infrared modifications to
gravity
might arise in a braneworld model
where our Universe is a 4D brane embedded in a 5D bulk.

The two branches of cosmological solutions in the DGP model have distinct properties.
In the so-called self-accelerating branch,
late-time acceleration of the Universe occurs
without the need of a cosmological constant~\cite{deffayet:01}.   Unfortunately
without a cosmological constant, the self-accelerating branch predicts cosmological
observables that are now in substantial conflict with the data (e.g.~\cite{fairbairn:06,maartens:06,fang:08}).  Moreover,
the linearized theory implies the presence of ghost degrees of freedom (e.g.~\cite{luty:03,charmousis:06}).
The former problem can be alleviated with the restoration of a cosmological constant or
brane tension. A definitive assessment of the latter problem awaits nonlinear solutions
\cite{dvali:06,koyama:07}.  On the second or normal branch, self-acceleration does not occur but interestingly
phantom effective equations of state with $p/\rho<-1$ can be achieved without ghosts
with the help of brane tension \cite{sahni:02}.   In both cases, brane tension is required but substantial modifications
to large-scale gravitational dynamics can still persist.

In this paper, we conduct a Markov chain Monte Carlo (MCMC)
 study of both branches of the DGP model using data from cosmic microwave background (CMB) anisotropies, supernovae distances, and the Hubble constant. For observables in the linear regime, we adopt the parametrized post-Friedmann (PPF) framework~\cite{hu:07, hu:08} and its implementation into a standard Einstein-Boltzmann linear theory solver~\cite{fang:08, fang:08b} for the theoretical predictions.
 This framework allows us to include information from the near horizon scales which are
 crucial for assessing the viability of the self-accelerating branch.  We also utilize
 information from the cross correlation between high-redshift galaxies and the CMB which
 has been proposed as an interesting test of both the self-accelerating and normal
 branches \cite{song:06,song:07,giannantonio:08}.

In Sec.~\ref{sec:theory}, we review the theory of the normal and self-accelerating branches of DGP gravity and their approximation through the PPF formalism. We present the results of our MCMC study in Sec.~\ref{sec:constraints} and discuss them in Sec.~\ref{sec:discussion}. Finally, the details about the modifications to the ISWWLL code~\cite{ho:08, hirata:08} used for the galaxy-ISW cross-correlation observations are specified in the Appendix.

\section{Normal and Self-accelerating Branches}\label{sec:theory}

In the DGP model~\cite{dvali:00} our Universe is a (3+1)-brane
embedded in a 5D Minkowski space described by the action
\begin{eqnarray}
S & = & -\frac{1}{2\kappa^2}\int d^5x \sqrt{-\hat{g}} \hat{R} -\frac{1}{2\mu^2}\int d^4x \sqrt{-\tilde{g}} \tilde{R} \nonumber \\
& & + \int d^4 x \sqrt{-\tilde{g}}L_T,
\end{eqnarray}
where 5D quantities are denoted by hats and 4D quantities are
denoted by tildes.
Matter fields, including a cosmological constant or brane tension and represented by $L_T$, are confined to the brane while only gravity extends to the full 5D bulk.   We assume that there is no bulk tension.
The constants $\kappa^2$ and $\mu^2$ are proportional to the inverse Planck masses in the bulk and brane, respectively.

Gravity on the brane is consequently modified at
large scales. In particular, the crossover distance $r_c = \kappa^2/2\mu^2$ governs the transition from 5D to 4D scalar-tensor gravity.  On scales smaller than the Vainshtein radius
$r_* = (r_c^2 r_g)^{1/3}$, nonlinear interactions return gravity to general relativity around
a point mass with Schwarzschild radius $r_g$.     In the following sections we
describe the evolution of the background and linear density perturbations in the two
branches of cosmological solutions.

\subsection{Background expansion}

Variation of the action yields the modified Einstein equations on the brane which reduce
to the modified Friedmann equation in a homogeneous and isotropic metric
\begin{equation}
H^2 - \frac{\sigma}{r_c} \sqrt{H^2+\frac{K}{a^2}} = \frac{\mu^2}{3} \sum_i \rho_i - \frac{K}{a^2},
\label{eq:friedmann}
\end{equation}
where $H=\dot a/a$ is the Hubble parameter,
$K$ is the spatial curvature, $a$ is the scale factor, and $\rho_i$ are the energy densities of the various components on the brane.
Here $\sigma = \pm 1$ and designates the branch of the cosmological solutions.

For $\sigma=+1$,  late-time acceleration occurs even without a cosmological constant $\Lambda$~\cite{deffayet:01} and so this branch is referred to as self-accelerating DGP (sDGP).
In order to separate tests of gravity from explanations of acceleration, we will also study the sDGP branch supplemented by a nonvanishing $\rho_{\Lambda}$ which we denote as sDGP+$\Lambda$ where confusion might arise.
 For $\sigma=-1$,
DGP modifications slow the expansion rate and the branch is referred to as normal DGP (nDGP). Here, a cosmological constant is required in order to achieve late-time acceleration.

%It is convenient to rescale the dimensionful quantities to the present Hubble scale $H_0$.
With the usual definitions
 $\Omega_i=\mu^2\rho_i(a=1)/H_0^2$ and $\Omega_K = -K/H_0^2$, the modified Friedmann
 equation becomes
  \begin{eqnarray}
% \left( \frac{H}{H_0} \right)^2 & = & \left( \sqrt{\Omega_m a^{-3} + \Omega_r a^{-4} + \Omega_{\Lambda} + \Omega_{r_c}} + \sigma \sqrt{\Omega_{r_c}} \right)^2 \nonumber\\
% & & + {\Omega_K}{a^{-2}},
 \left( \frac{H}{H_0} \right)^2 & = & \left( \sqrt{\frac{\Omega_m}{a^3} + \frac{\Omega_r}{a^4} + \Omega_{\Lambda} + \Omega_{r_c}} + \sigma \sqrt{\Omega_{r_c}} \right)^2 \nonumber\\
 & & + \frac{\Omega_K}{a^2},
\label{eq:hubble_parameter}
\end{eqnarray}
where we have assumed that the energy density components include nonrelativistic matter,
radiation, and possibly a cosmological constant.
Here
\begin{equation}
\sqrt{\Omega_{r_c}} \equiv \frac{1}{2H_0r_c} = \sigma \frac{\Omega_{\rm DGP}}{2\sqrt{1-\Omega_K}},
\label{eq:omegarc}
\end{equation}
where
\begin{eqnarray}
\Omega_{\rm DGP} &=& 1-\Omega_m-\Omega_r-\Omega_{\Lambda}-\Omega_K %\nonumber\\
% & =  & 2\sigma\sqrt{\Omega_{r_c}(1-\Omega_K)}
\end{eqnarray}
represents the effective contribution of the DGP modification to the energy density assuming
the ordinary Friedmann equation.
Specifically,
\begin{equation}
\rho_{\rm DGP} \equiv \frac{3}{\mu^2} \left( H^2 + \frac{K}{a^2} \right) - \sum_i \rho_i.
\label{eq:rho_e}
\end{equation}
As with any real energy density component, it obeys the conservation equation
\begin{equation}
\rho_{\rm DGP}'=-3(1+w_{\rm DGP})\rho_{\rm DGP}.
\label{eq:fluid_equation}
\end{equation}
Using Eqs.~(\ref{eq:friedmann}) and (\ref{eq:fluid_equation}), we derive
\begin{equation}
1+w_{\rm DGP} = \frac{ \frac{\mu^2}{3} \sum_i (1+w_i) \rho_i }{ H^2 + \frac{K}{a^2} + \frac{\mu^2}{3} \sum_i \rho_i }.
\label{eq:w_DGP}
\end{equation}

For cases with a cosmological constant it is also useful to define the total
effective dark energy
\begin{equation}
\rho_e = \rho_{\rm DGP} + \rho_\Lambda
\end{equation}
and its equation of state
\begin{equation}
1+w_e = (1+w_{\rm DGP})\frac{\rho_{\rm DGP}}{\rho_{\rm DGP} + \rho_\Lambda}  \,.
\end{equation}
In nDGP this quantity diverges when the DGP and $\Lambda$ density terms
are equal and opposite at which point the value of $1+w_e$ switches sign.
In particular, its value today is given by
\begin{equation}
w_0 = -\frac{1-\Omega_K}{1-\Omega_K-\Omega_m} \frac{1-\Omega_K-\Omega_m+\Omega_{\Lambda}}{1-\Omega_K+\Omega_m+\Omega_{\Lambda}},
\label{eq:w0}
\end{equation}
where we have neglected the small radiation contribution.
With realistic cosmological parameters $w_0> -1$ and $w_0< -1$ for sDGP and nDGP, respectively, with
$w_0=-1$ being the limit of cosmological constant domination in either case.

\subsection{PPF linear theory}
\label{sec:ppf}

Unlike the background,
the evolution of density and metric perturbations on the brane require solutions for the bulk metric
equations.
The parametrized post-Friedmann framework was introduced in Refs.~\cite{hu:07, hu:08} to
encapsulate these effects in an effective 3+1 description.  Fits to the bulk calculation in
sDGP without $\Lambda$ or curvature from \cite{sawicki:06} were given in \cite{hu:07} and
incorporated into the linear theory code CAMB \cite{lewis:00} in \cite{fang:08}.   We extrapolate
these fits to cases with $\Lambda$ and curvature here though we caution the reader that these have not
been tested by explicit bulk calculations.   For nDGP, we utilize a description from
\cite{seahra:09} based on bulk calculations from \cite{song:07} and \cite{cardoso:08} with $\Lambda$
but no curvature.  We again extrapolate these results for spatial curvature.  The errors
induced by these extrapolations are controlled given the well-defined limits of $\Lambda$
domination and the small dynamical effects of curvature in the regime we consider.

 Given the expansion history, the PPF framework is defined by three functions and one parameter.  From these quantities, the dynamics are determined by conservation of
 energy and momentum and the Bianchi identities.     The defining quantities are
 $g(a,k)$ which quantifies the effective anisotropic stress of the modifications and distinguishes
 the two gravitational potentials,
 $f_\zeta(a)$, which defines the relationship between the matter and the metric on
 superhorizon scales, and $f_G(a)$, which defines it in the linearized Newtonian regime.  
 The additional parameter defines the transition scale between the superhorizon and
 Newtonian behaviors.

 More specifically, 
\begin{equation}
g(a,k) \equiv \frac{\Phi+\Psi}{\Phi-\Psi},
\end{equation}
where the scalar linear perturbations are represented in longitudinal gauge
\begin{equation}
ds^2 = -(1+2\Psi)dt^2 + a^2 (1+2\Phi)dx^2,
\end{equation}
where $dx^2$ is the unperturbed spatial line element with constant curvature $K$.
In the quasistatic high $k$ limit, the DGP model predicts
\begin{equation}
%g_{\rm QS} = -\frac{1}{3} \left[ 1 - \sigma \frac{2 H r_c }{\sqrt{1-\Omega_K(a)}} \left(1+\frac{H'}{3H}-\frac{2}{3} \Omega_K(a) \right) \right]^{-1},
g_{\rm QS} = -\frac{1}{3} \left[ 1 -  \frac{2\sigma H r_c }{\sqrt{1-\Omega_K(a)}} \left(1+\frac{H'}{3H}-\frac{2}{3} \Omega_K(a) \right) \right]^{-1},
\label{eq:gQS}
\end{equation}
where $\Omega_K(a) = H_0^2\Omega_K/H^2a^2$.
On superhorizon scales, we take for sDGP \cite{hu:07} 
\begin{eqnarray}
g_{\rm SH, sDGP}(a) & = & \frac{9}{8 H r_c \sqrt{1-\Omega_K(a)} - 1}\\
& & \times \left( 1 + \frac{0.51}{H r_c \sqrt{1-\Omega_K(a)} - 1.08} \right). \nonumber
\label{eq:gSHsDGP}
\end{eqnarray}
We exclude models  $\sqrt{1-\Omega_K}H_0 r_c > 1.08$ from consideration as they
are not within the domain of applicability of the fit nor are they cosmologically
viable.
For nDGP we take \cite{seahra:09} ({\it cf.} \cite{afshordi:08})
\begin{equation}
g_{\rm SH, nDGP}(a) = -\frac{1}{2H r_c \sqrt{1-\Omega_K(a)} + 1}.
\label{eq:gSHnDGP}
\end{equation}

The corrections for curvature have not been verified by a bulk calculation
for the superhorizon cases. For the curvatures that we will consider the total
impact is small as can be verified by omitting the correction.  We expect
therefore that corrections on the correction have negligible impact.

At intermediate scales, $g$ is fitted by the interpolating function
\begin{equation}
g(a,k) = \frac{g_{\rm SH}+g_{\rm QS}(c_g k_H)^{n_g}}{1+(c_g k_H)^{n_g}},
\end{equation}
where $k_{H}=k/aH$, $c_g=0.14$ for sDGP and $c_g=0.4$ for nDGP, respectively. Furthermore, we set $n_g = 3$.

The function $f_{\zeta}(\ln a)$ relates the metric to the density at superhorizon scales and is well described by $f_{\zeta}(\ln a) = 0.4 g_{\rm SH}(\ln a)$. In the quasistatic regime, the analogous
relationship between $\Phi-\Psi$ and the density is the Poisson equation and that is unmodified from ordinary gravity for
DGP. Hence $f_G(\ln a) = 0$.

Finally the parameter $c_{\Gamma}$ relates the transition scale in the dynamical equations
from superhorizon to quasistatic behavior. For sDGP we take $c_{\Gamma}=1$ following
\cite{hu:07} and we employ this value for cases that include $\Lambda$.
In nDGP, $c_{\Gamma} \sim 0.15$ from \cite{seahra:09}
implying a delayed approach to quasistatic behavior.

\section{Constraints on the Models}\label{sec:constraints}

We will use a variety of cosmological data sets to constrain the two branches of the DGP models.
First we use the CMB anisotropy data from the five-year Wilkinson Microwave Anisotropy Probe (WMAP)~\cite{WMAP:08}, the Arcminute Cosmology Bolometer Array Receiver (ACBAR)~\cite{ACBAR:07}, the Cosmic Background Imager (CBI)~\cite{CBI:04}, and the Very Small Array (VSA)~\cite{VSA:03}. Next we employ data from the Supernovae Legacy Survey (SNLS)~\cite{SNLS:06} and the measurement of the Hubble constant from the Supernovae and $H_0$ for the Equation of State (SHOES)~\cite{SHOES:09} program. Finally we take galaxy-ISW (gISW) correlation observations using the likelihood code of ~\cite{ho:08, hirata:08}. We quote results with and without the gISW constraint to highlight its impact on the
results.

In Sec.~\ref{sec:predictions} we discuss the predictions for these observables in the two
branches of the DGP model.  In Secs.~\ref{sec:flat} and \ref{sec:nonflat} we present the results
of a MCMC-likelihood analysis for flat and nonflat universes,
respectively.
The MCMC analysis is conducted with the publicly available CosmoMC~\cite{lewis:02} package.

\subsection{Model predictions}
\label{sec:predictions}

In this section we illustrate model predictions of the
various cosmological observables we use in the
constraints.   We chose the parameters of the various models that highlight results
from the MCMC analysis.

At high redshifts the DGP modifications become negligible on either
branch [see Eq.~(\ref{eq:hubble_parameter})], and so we
choose a parametrization that separates high-redshift and low-redshift constraints.
Specifically we take 6 high-redshift parameters: the physical baryon and cold dark
matter energy density
$\Omega_bh^2$ and $\Omega_ch^2$, the
ratio of sound horizon to angular diameter distance at recombination multiplied by a factor of 100
$\theta$, the optical depth to reionization $\tau$, the scalar tilt $n_s$, and amplitude
$A_s$ at $k_*=0.002~\textrm{Mpc}^{-1}$.

The low-redshift parameters differ in the various classes of models.   For flat $\Lambda$CDM
and sDGP without $\Lambda$ there are no additional degrees of freedom.  Note that $\theta$ carries information on $H_0$.  For flat sDGP$+\Lambda$ and nDGP, $\Omega_{\Lambda}$
is an extra degree of freedom.   For the open versions of all models $\Omega_K$ is the
final degree of freedom.

For $\Lambda$CDM and sDGP we illustrate predictions from the nonflat maximum likelihood
models found in the next section (see Tables~\ref{tab:res_k_lcdm} and \ref{tab:res_k_sdgp}).
 Since the large-scale behavior of nDGP is new to this work, we highlight the dependence of
 observables
 on
$\Omega_{\Lambda}$ and $\Omega_K$ while keeping the high-redshift cosmological parameters fixed (see Table~\ref{tab:ndgp_models}). Note in the $r_c \rightarrow 0$ limit where
$\Omega_{r_c}=0$,
both nDGP and sDGP+$\Lambda$ become $\Lambda$CDM.
We therefore choose to illustrate the maximum likelihood
 sDGP model with $\Lambda=0$.

\begin{table}[ht]
\centering
\begin{tabular}{|c|c|c|c|c|c|c|c|}
\hline
nDGP & A & B & C & D & E & F & G \\
\hline
$\Omega_{\Lambda}$ & 0.77  & 1.00  & 1.25  & 1.50  & 1.50   & 1.25   & 1.00   \\
$\Omega_K$         & --    & --    & --    & --    & -0.025 & -0.015 & -0.010 \\
$\Omega_{r_c}$     & 0.000 & 0.012 & 0.049 & 0.114 & 0.132  &  0.057 & 0.015  \\
$H_0$              & 73    & 77    & 82    & 86    & 71     & 72     & 71     \\
\hline
\end{tabular}
\caption{Different choices of nDGP models for illustration. Note
that nDGP-A is the best-fit (with gISW) flat nDGP model, corresponding to $\Lambda$CDM. Other
chain parameters are fixed to values in Table~\ref{tab:res_flat_ndgp}.}
\label{tab:ndgp_models}
\end{table}

\subsubsection{Cosmic microwave background}

\begin{figure*}
\resizebox{\hsize}{!}{\includegraphics{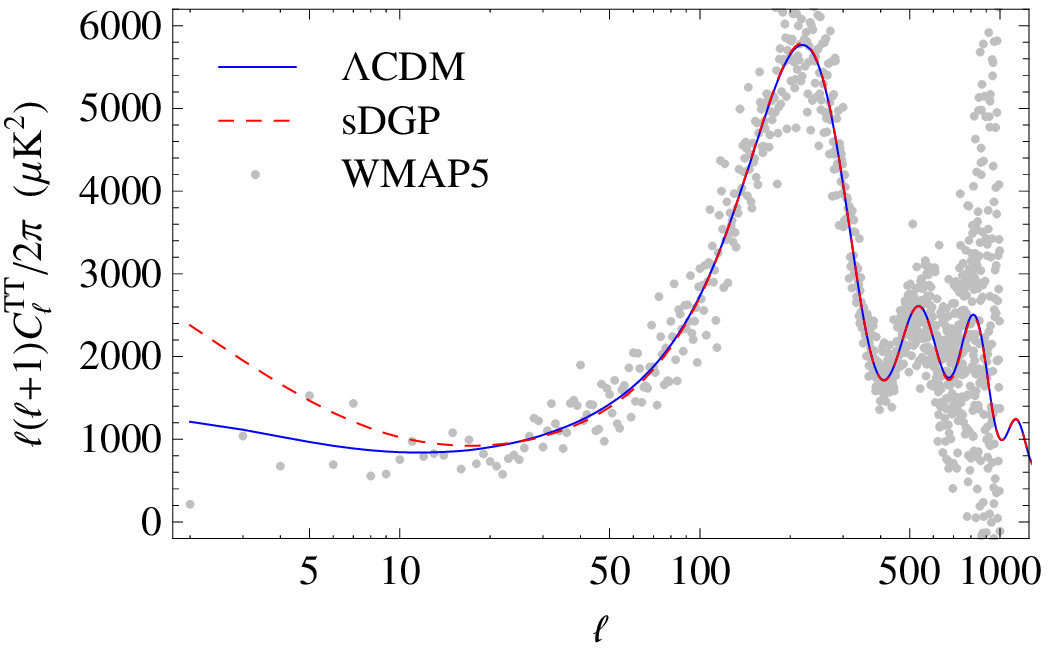}\includegraphics{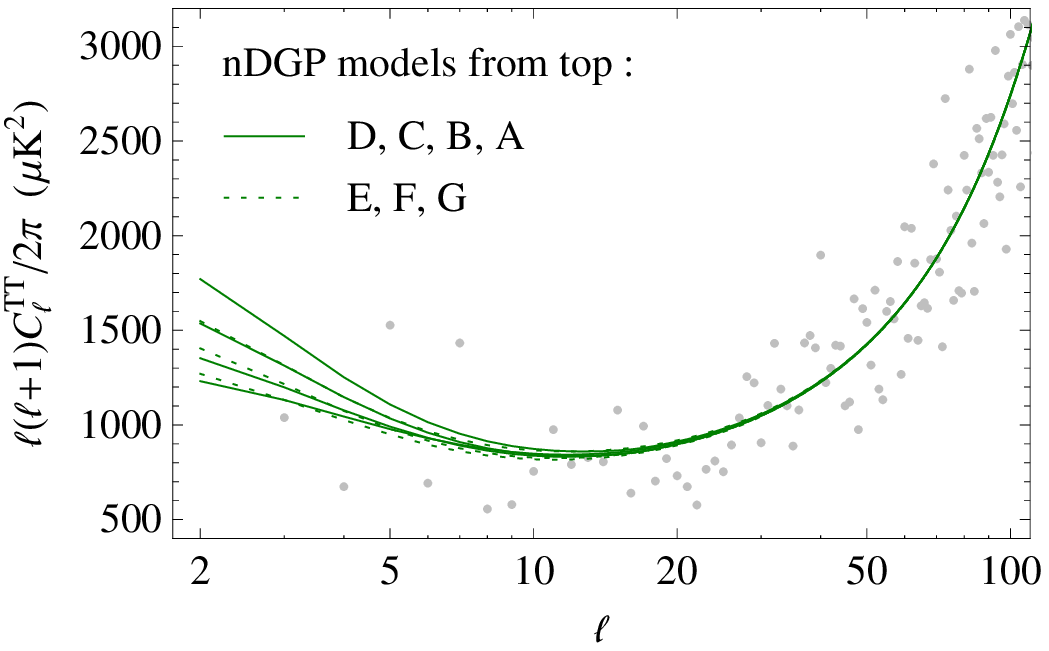}}
\caption{Best-fit CMB temperature anisotropy power spectrum for $\Lambda$CDM and sDGP (left panel). Examples of nDGP models (right panel) illustrate the degeneracy between $\Omega_{\Lambda}$ and $\Omega_K$ corresponding to models in Table~\ref{tab:ndgp_models}.}
\label{fig:sDGPL}
\end{figure*}

The CMB probes the geometry of the background expansion
 as well as the formation of large-scale structure.   The latter manifests itself on the largest scales
 through the integrated Sachs-Wolfe (ISW) effect  from the evolution of the gravitational potential.
 To predict these effects we implement the PPF modifications from Sec.~\ref{sec:ppf}.
The incorporation of the PPF formalism into a standard Einstein-Boltzmann linear theory solver yields an efficient way to obtain predictions of the DGP model for the CMB. We utilize the PPF modifications to CAMB~\cite{lewis:00} implemented in Refs.~\cite{fang:08, fang:08b}, which we can apply directly for sDGP and figure as a starting point for the implementation of nDGP and sDGP+$\Lambda$. In Fig.~\ref{fig:sDGPL}, we plot the CMB temperature anisotropy power spectrum with respect to angular multipole $\ell$ for the best-fit models of $\Lambda$CDM and sDGP, as well as the nDGP parameter choices given in Table~\ref{tab:ndgp_models}.

 Relative to $\Lambda$CDM,
the growth of structure is suppressed in the sDGP model, yielding an
ISW enhancement at the lowest multipoles.   This enhancement is
sufficiently large to bring the sDGP model without $\Lambda$ into
serious conflict with the joint data \cite{fang:08}. The opposite
effects occurs in the nDGP model and lead to predictions that are
compatible with CMB data.  Here raising $\Omega_\Lambda$ at fixed
$\Omega_K$ enhances the low multipoles through the ISW effect.
However, compensating effects from curvature can lead to parameter
degeneracies.

At high redshifts the contribution of $\Omega_{r_c}$ to the Hubble parameter, Eq.~(\ref{eq:hubble_parameter}), becomes negligible in either branch. The CMB acoustic peaks can therefore be utilized as usual to infer constraints on the high-redshift parameters,
in particular, the physical energy densities of baryonic matter and dark matter as well as the
angular diameter distance to recombination.

\subsubsection{Distances to the supernovae and $H_0$}

The comparison of the magnitudes of high-redshift to low-redshift
supernovae yields a relative distance measure. Theoretical
predictions for the distance modulus are related to the luminosity
distance, $d_L(z) = (1+z) r(z)$, where $r(z)$ is the comoving
angular diameter distance defined by
\begin{equation}
r(z) = \left\{
\begin{array}{ll}
\sin\left[H_0 \sqrt{-\Omega_K} \chi(z)\right]/H_0\sqrt{|\Omega_K|}, & \Omega_K < 0, \\
\chi(z), & \Omega_K = 0, \\
\sinh\left[H_0 \sqrt{\Omega_K} \chi(z)\right]/H_0\sqrt{|\Omega_K|}, & \Omega_K > 0,
\end{array}
\right.
\end{equation}
where the comoving radial distance $\chi$ is
\begin{equation}
\chi(z) = \int_0^z \frac{dz'}{H(z')}.
\end{equation}
The supernovae magnitudes, once standardized, are related
to the distance by
\begin{equation}
m \equiv \mu + M = 5 \log_{10}{d_L}+M + 25,
\end{equation}
where $d_L$ is in units of Mpc. The unknown absolute magnitude $M$ of the supernovae is
a nuisance parameter in the fit and is degenerate with $H_0$.  Hence supernovae
measure relative distances within the set.

\begin{figure}
\resizebox{\hsize}{!}{\includegraphics{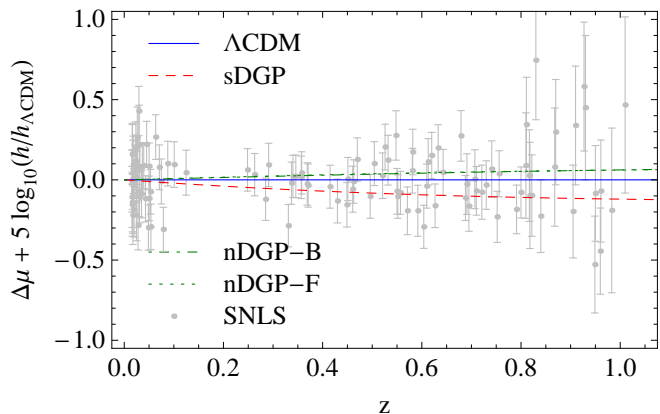}}
\caption{Best-fit distance modulus for sDGP, as well as the overlapping predictions for nDGP-B and nDGP-F with respect to $\Lambda$CDM.}
\label{fig:dL}
\end{figure}

In Fig.~\ref{fig:dL}, we plot the predictions for the distance modulus for the SNLS data in sDGP gravity, nDGP-B, nDGP-F, and in the $\Lambda$CDM model.

The acoustic peaks in the CMB and the measurement of the local Hubble constant additionally provide absolute distance probes which complement the relative distance measure of the supernovae.  
For the Hubble constant, we take the SHOES measurement of $H_0 = 74.2\pm 3.6$
km s$^{-1}$ Mpc$^{-1}$ which employs Cepheid measurements to link the low-redshift supernovae to the distance 
scale established by the maser galaxy NGC 4258.

\subsubsection{Galaxy-ISW cross correlations}

The correlation between galaxy number densities and the CMB
anisotropies can be used to isolate the ISW effect in the CMB. The
enhanced ISW effect exhibited in the sDGP model without $\Lambda$
 leaves a strong imprint on the large scales of the CMB temperature anisotropy. As pointed out by Song et al.~\cite{song:06}, an interesting consequence of this is a considerable correlation of high-redshift galaxies with the CMB.

For nDGP gravity, whereas the ISW effect does not exhibit a substantial impact on the largest scales in the CMB,  useful signatures remain in the correlations with galaxies that can
break parameter degeneracies~\cite{giannantonio:08}.

We evaluate the gISW cross correlations in the  Limber and quasistatic approximation, as it is done in the ISWWLL code~\cite{ho:08, hirata:08} used for the data analysis. Therefore, we write
\begin{eqnarray}
C_{\ell}^{g_jT} & \simeq & \frac{3 \Omega_m H_0^2 T_{\rm CMB}}{(\ell+1/2)^2} \int dz \: f_j(z) H(z) D(z) \nonumber\\
& & \times \frac{d}{dz} [ D(z) (1+z)] P \left( \frac{\ell+1/2}{\chi(z)} \right).
\label{eq:ClgT}
\end{eqnarray}
Here, $D(z)$ is the linear growth rate in the quasistatic regime defined by $\Delta_m(k,z) = \Delta_m(k,0) D(z)/D(0)$, where $\Delta_m(k,z)$ is the matter density perturbation. $P(k)$ is
the matter power spectrum today.

The approximations in Eq.~(\ref{eq:ClgT}) become accurate at the percent level for $\ell\gtrsim10$.
This condition is satisfied by about 90\% of the total 42 data points that are used in the ISWWLL code. We discuss details about the approximations and the data in the Appendix. The data are divided into nine galaxy sample bins $j$, i.e., 2MASS0-3, LRG0-1, QSO0-1, and NVSS. The function $f_j(z)$ relates the matter density to the observed projected galaxy overdensity with $f_j(z) = b_j(z) \Pi_j(z)$ in the absence of magnification bias. $\Pi_j (z)$ is the redshift distribution of the galaxies and the bias factor $b_j(z)$ is assumed independent of scale, but dependent on redshift. The code determines $f_j(z)$, among other things, from fitting autopower spectra and cross-power spectra between the samples.

\begin{figure*}
\resizebox{\hsize}{!}{\includegraphics{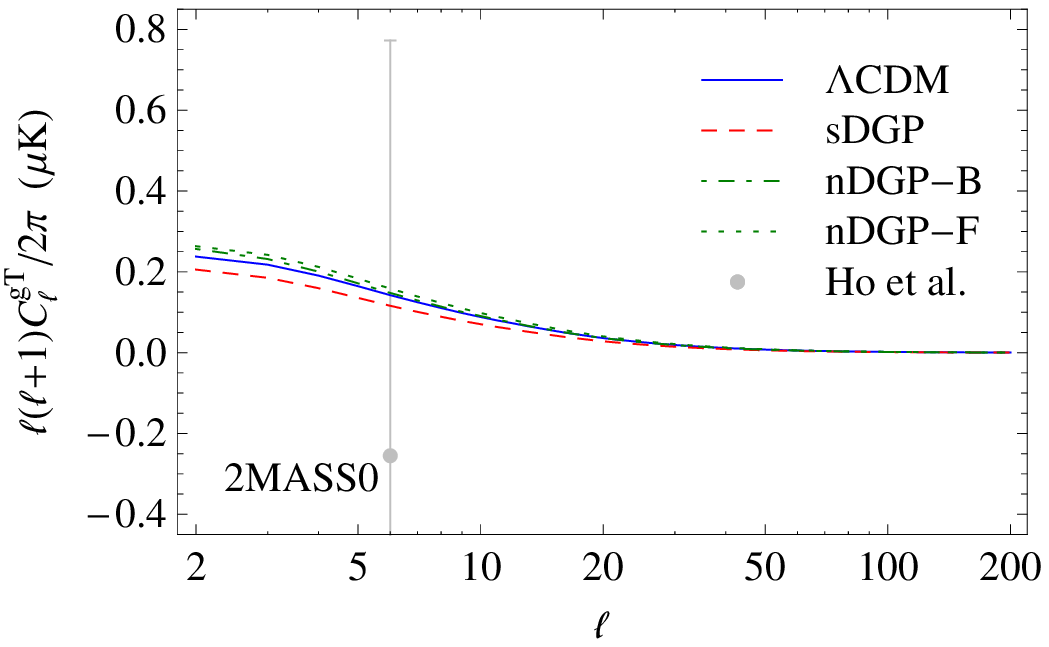}\includegraphics{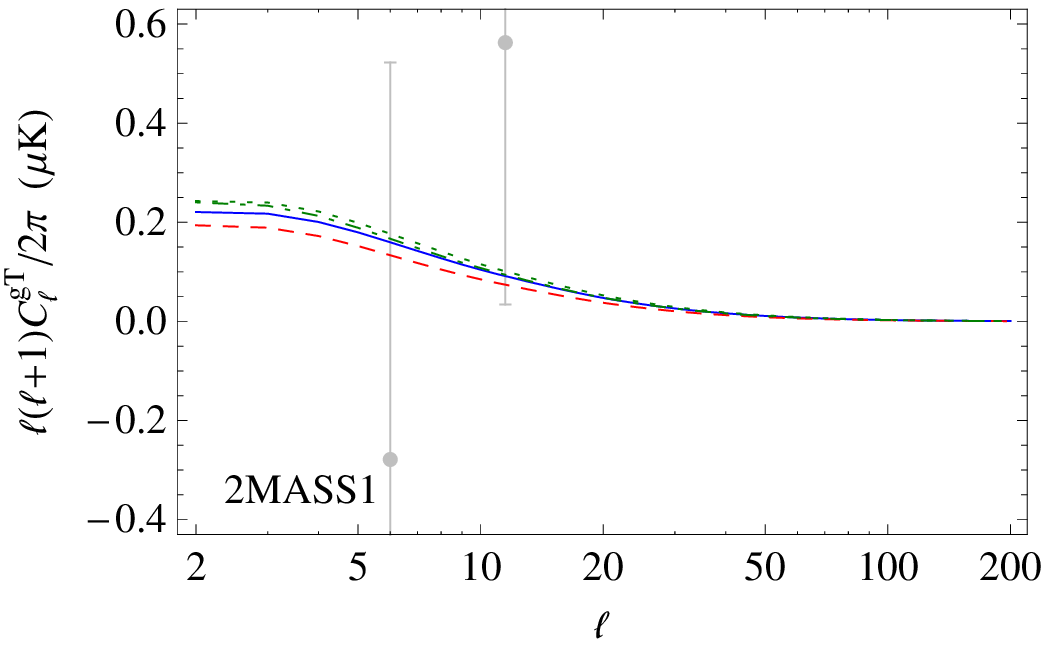}\includegraphics{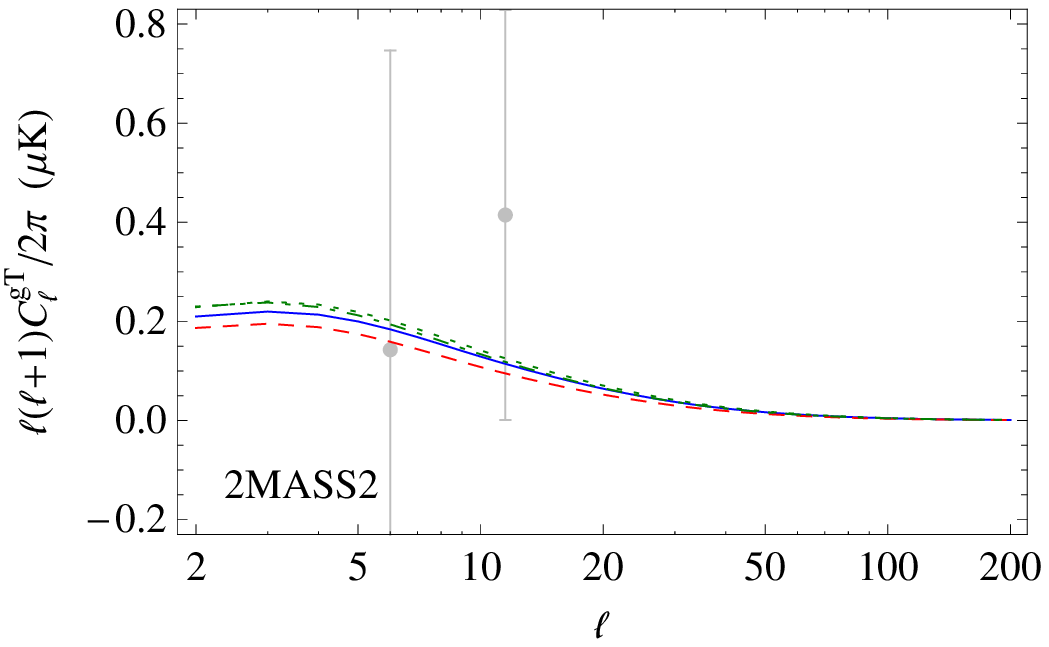}}
\resizebox{\hsize}{!}{\includegraphics{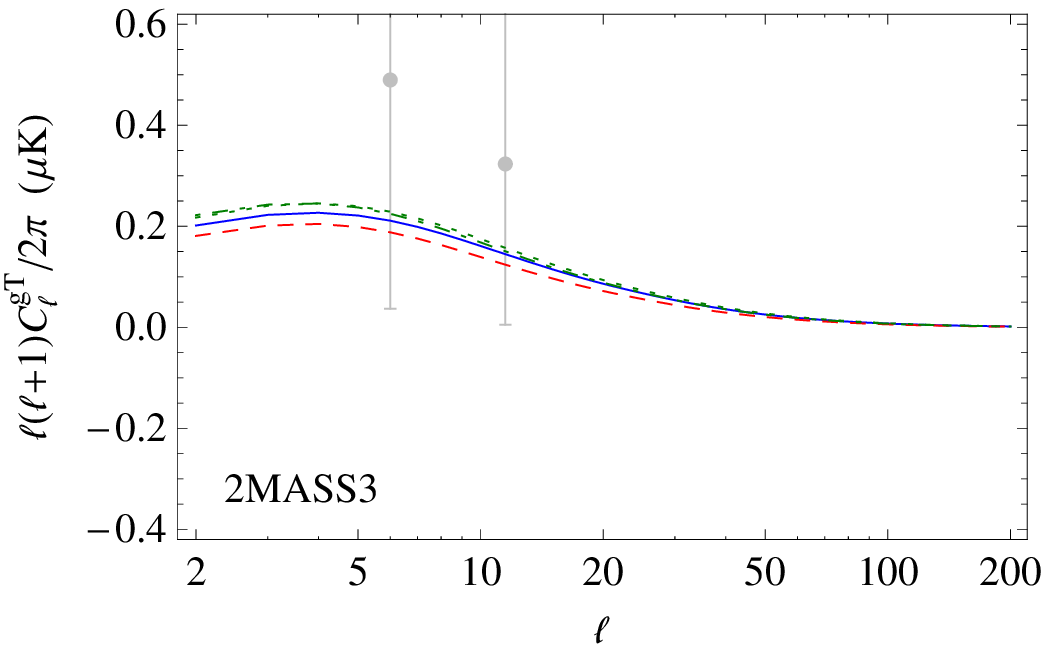}\includegraphics{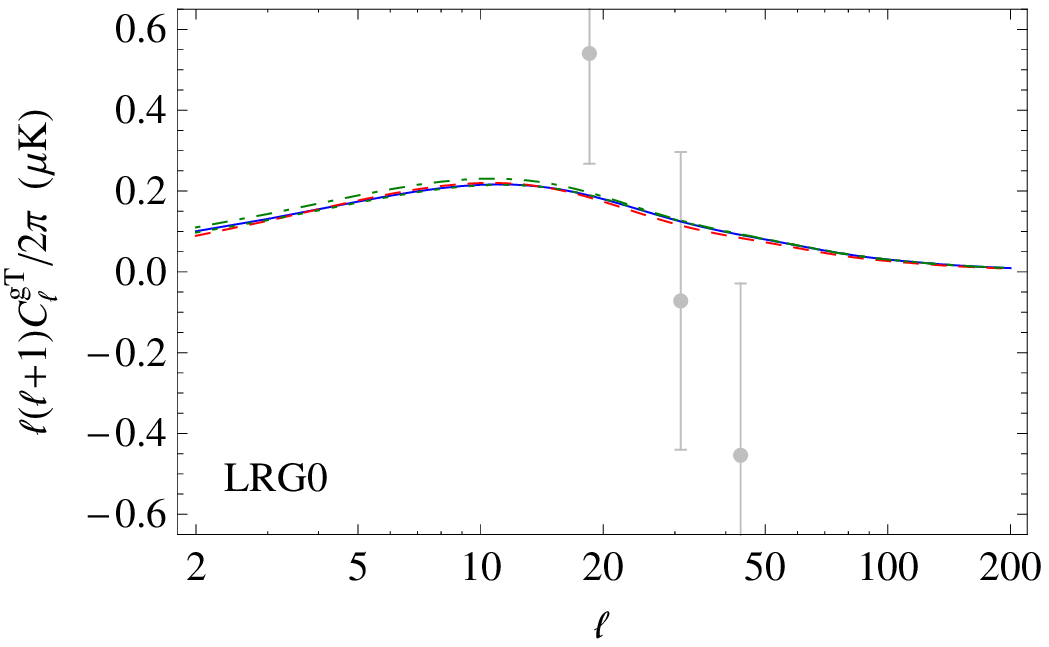}\includegraphics{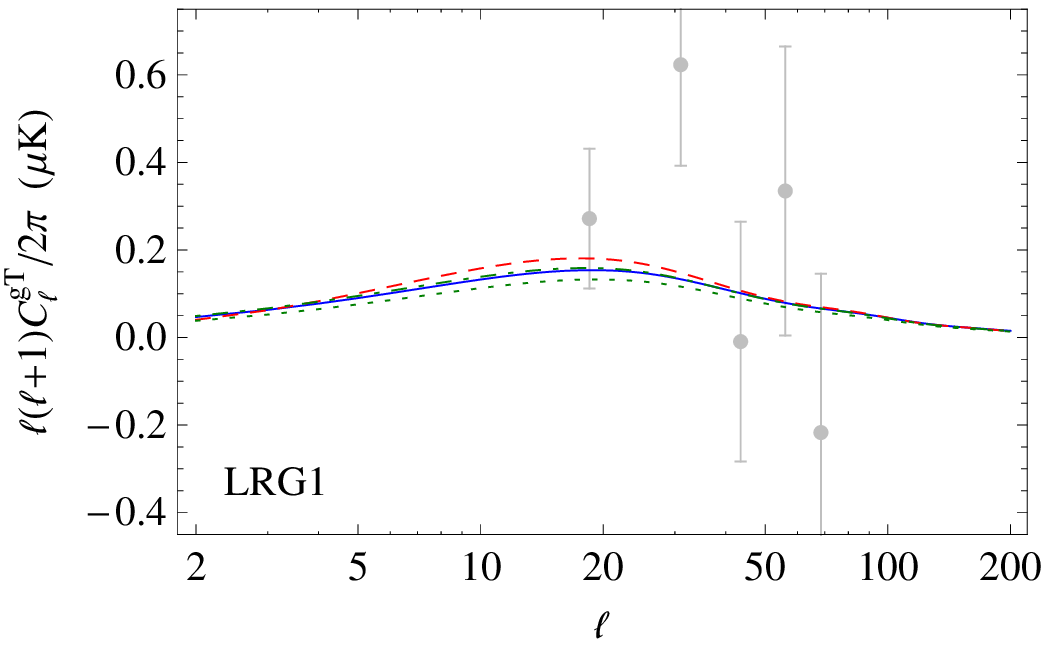}}
\resizebox{\hsize}{!}{\includegraphics{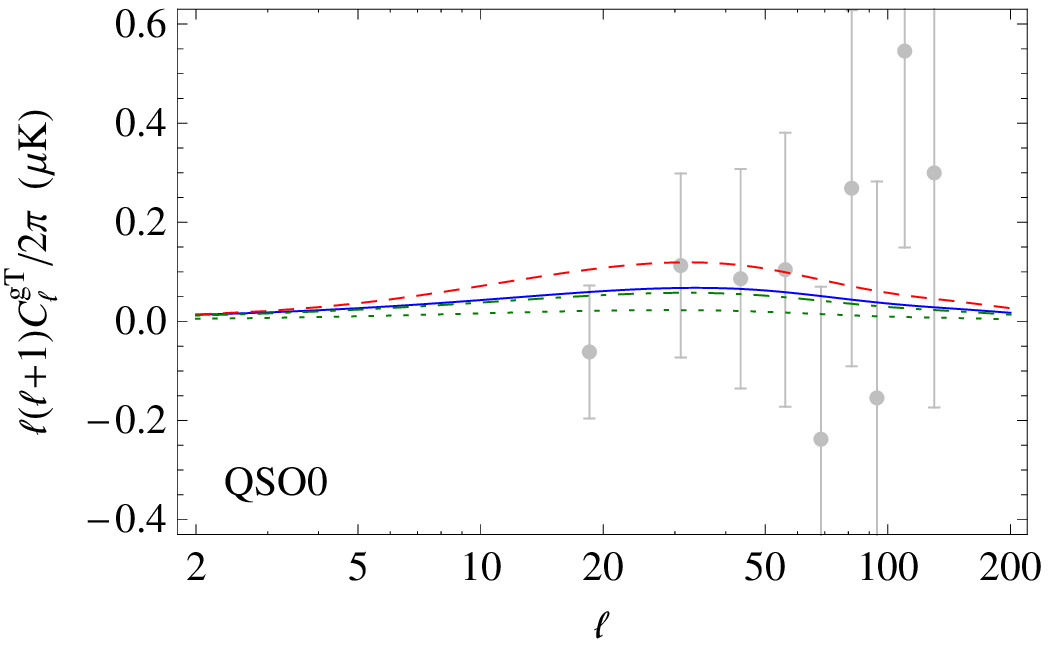}\includegraphics{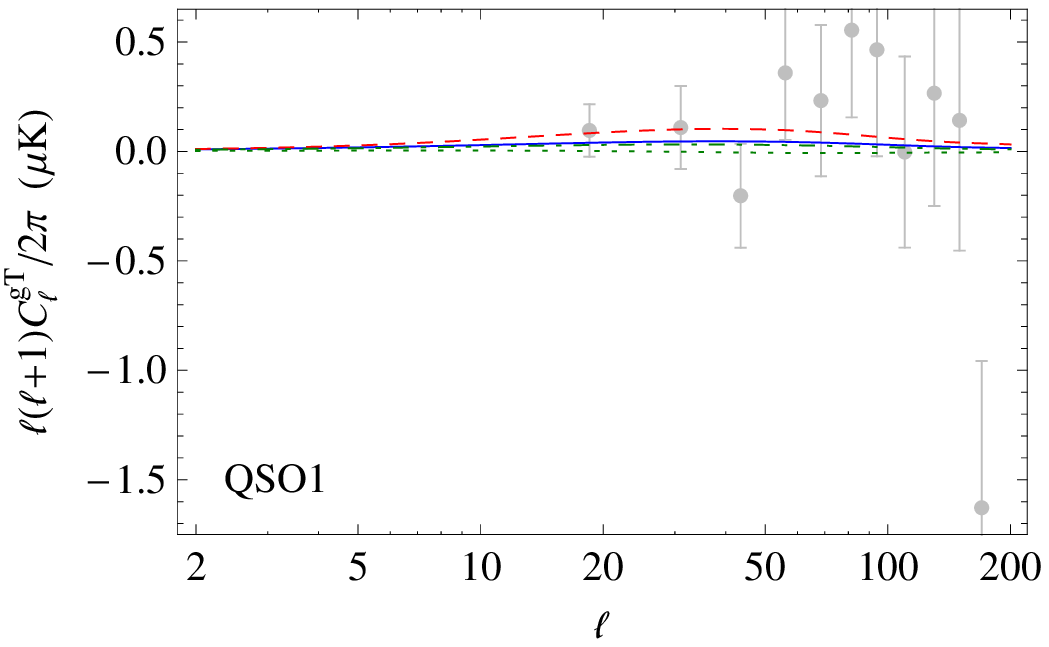}\includegraphics{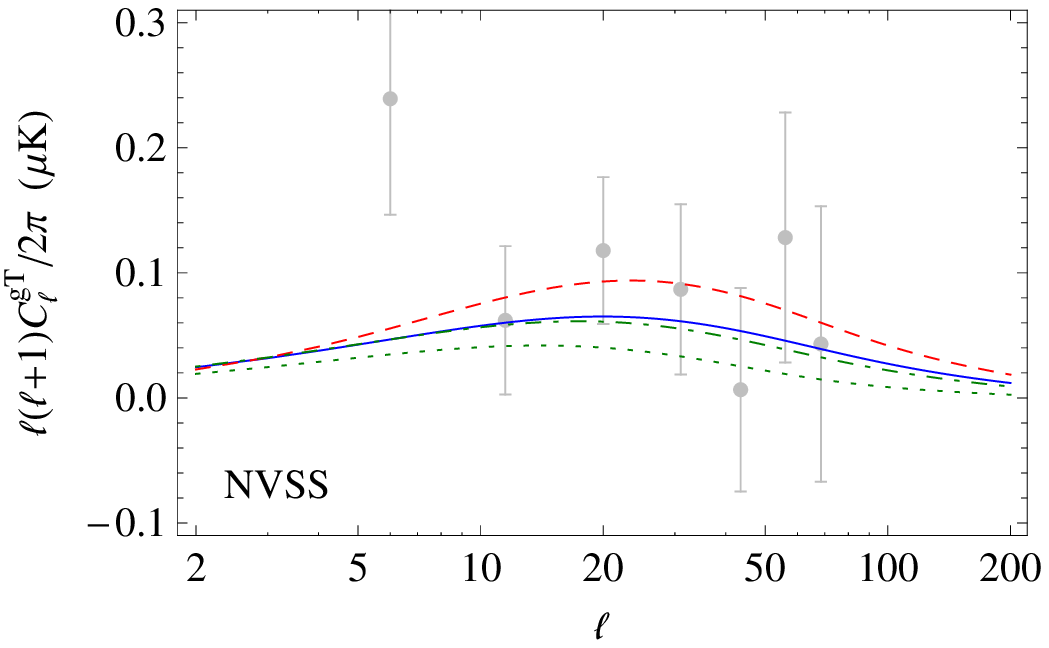}}
\caption{Best-fit $\Lambda$CDM and sDGP galaxy-ISW cross correlations for the different galaxy samples, roughly ordered in increasing effective, bias-weighted, redshift. Note the distinct predictions for the previously degenerate nDGP-B and nDGP-F models.}
\label{fig:ISW}
\end{figure*}

We modify the above calculations in the ISWWLL code with the appropriate DGP quantities such that the correct predictions for the crosscorrelations are obtained. We refer to the Appendix for details. The predictions for the best-fit values, combining all data, of $\Lambda$CDM and sDGP for the different samples are shown in Fig.~\ref{fig:ISW}. We also plot the curves for nDGP-B and nDGP-F to illustrate the breaking of the degeneracy between $\Omega_{\Lambda}$ and $\Omega_K$. Notice that the model with larger curvature has
reduced correlation especially at high redshift.  We shall see that models with significantly
larger curvature can be excluded by the gISW data.

\subsection{Flat universe constraints}
\label{sec:flat}

We begin by studying a universe without spatial curvature, where the basic cosmological parameter set is $P = \{ \Omega_bh^2, \Omega_ch^2, \theta, \tau, n_s, \ln[10^{10}A_s] \}$.  We implement the following flat priors on them: $\Omega_bh^2\in(0.01,0.1)$, $\Omega_ch^2\in(0.05,0.99)$, $\theta\in(0.5,10)$, $\tau \in (0.01,0.8)$,
$n_s\in(0.5,1.5)$, and $\ln[10^{10}A_s]\in(2.7,4)$. For nDGP and sDGP with nonvanishing $\Lambda$, we use $\Omega_{\Lambda}\in(0.0,2.5)$.

\begin{table}[ht]
\resizebox{\hsize}{!}{
\centering
\begin{tabular}{|l|r@{}|c|c|r@{}|c|c|}
\hline
%\cline{1-1} \cline{3-4} \cline{6-7}
Parameters & & \multicolumn{2}{|c|}{$\Lambda$CDM} & & \multicolumn{2}{|c|}{$\Lambda$CDM (with gISW)} \\
\cline{1-1} \cline{3-4} \cline{6-7}
$100\Omega_b h^2$  & & $2.248\pm0.055$   & 2.240  & & $2.251\pm0.055$   & 2.258  \\
$\Omega_c h^2$     & & $0.1080\pm0.0043$ & 0.1072 & & $0.1075\pm0.0042$ & 0.1071 \\
$\theta$           & & $1.0410\pm0.0027$ & 1.0404 & & $1.0411\pm0.0027$ & 1.0417 \\
$\tau$             & & $0.086\pm0.017$   & 0.086  & & $0.087\pm0.017$   & 0.089  \\
$n_s$              & & $0.963\pm0.013$   & 0.961  & & $0.963\pm0.013$   & 0.965  \\
$\ln[10^{10}A_s]$  & & $3.176\pm0.041$   & 3.177  & & $3.174\pm0.041$   & 3.173  \\
\cline{3-4} \cline{6-7}
$\Omega_{\Lambda}$ & & $0.751\pm0.019$   & 0.754  & & $0.754\pm0.019$   & 0.758  \\
$\Omega_m$         & & $0.249\pm0.019$   & 0.246  & & $0.246\pm0.019$   & 0.242  \\
$H_0$              & & $72.6\pm1.8$      & 72.6   & & $72.8\pm1.8$      & 73.2   \\
\cline{1-1} \cline{3-4} \cline{6-7}
$-2\ln L$ & & \multicolumn{2}{|c|}{2834.29} & & \multicolumn{2}{|c|}{2867.99} \\
\cline{1-1} \cline{3-4} \cline{6-7}
\end{tabular}
}
\caption{Means,  standard deviations (left subdivision of columns), and best-fit values (right subdivision of columns) with likelihood for the flat $\Lambda$CDM model using data from WMAP, ACBAR, CBI, VSA, SNLS, and SHOES without (left column) and with the gISW data (right column).}
\label{tab:res_flat_lcdm}
\end{table}

\begin{table}[ht]
\resizebox{\hsize}{!}{
\centering
\begin{tabular}{|l|r@{}|c|c|r@{}|c|c|}
\hline
%\cline{1-1} \cline{3-4} \cline{6-7}
Parameters & & \multicolumn{2}{|c|}{sDGP} & & \multicolumn{2}{|c|}{sDGP (with gISW)} \\
\cline{1-1} \cline{3-4} \cline{6-7}
$100\Omega_b h^2$  & & $2.390\pm0.066$   & 2.393  & & $2.390\pm0.065$   & 2.376  \\
$\Omega_c h^2$     & & $0.0884\pm0.0042$ & 0.0873 & & $0.0889\pm0.0041$ & 0.0899 \\
$\theta$           & & $1.0448\pm0.0028$ & 1.0447 & & $1.0449\pm0.0028$ & 1.0452 \\
$\tau$             & & $0.105\pm0.021$   & 0.110  & & $0.105\pm0.021$   & 0.103  \\
$n_s$              & & $1.011\pm0.015$   & 1.013  & & $1.011\pm0.015$   & 1.007  \\
$\ln[10^{10}A_s]$  & & $3.001\pm0.045$   & 2.998  & & $3.003\pm0.044$   & 3.015  \\
\cline{3-4} \cline{6-7}
$\Omega_{r_c}$     & & $0.1410\pm0.0075$ & 0.1430 & & $0.1403\pm0.0075$ & 0.1384 \\
$\Omega_m$         & & $0.249\pm0.020$   & 0.244  & & $0.251\pm0.020$   & 0.256  \\
$H_0$              & & $67.2\pm1.7$      & 67.6   & & $67.1\pm1.7$      & 66.7   \\
\cline{1-1} \cline{3-4} \cline{6-7}
$-2\Delta\ln L$ & & \multicolumn{2}{|c|}{32.70} & & \multicolumn{2}{|c|}{33.06} \\
\cline{1-1} \cline{3-4} \cline{6-7}
\end{tabular}
}
\caption{Same as Table~\ref{tab:res_flat_lcdm}, but for the flat sDGP model. $-2\Delta\ln L$ is
quoted with respect to the maximum likelihood flat $\Lambda$CDM model.}
\label{tab:res_flat_sdgp}
\end{table}

\begin{table}[ht]
\resizebox{\hsize}{!}{
\centering
\begin{tabular}{|l|r@{}|c|c|r@{}|c|c|}
\hline
%\cline{1-1} \cline{3-4} \cline{6-7}
Parameters & & \multicolumn{2}{|c|}{sDGP+$\Lambda$} & & \multicolumn{2}{|c|}{sDGP+$\Lambda$ (with gISW)} \\
\cline{1-1} \cline{3-4} \cline{6-7}
$100\Omega_b h^2$  & & $2.265\pm0.058$   & 2.245  & & $2.265\pm0.058$   & 2.257  \\
$\Omega_c h^2$     & & $0.1050\pm0.0046$ & 0.1071 & & $0.1048\pm0.0046$ & 0.1070 \\
$\theta$           & & $1.0415\pm0.0028$ & 1.0405 & & $1.0415\pm0.0027$ & 1.0415 \\
$\tau$             & & $0.089\pm0.017$   & 0.080  & & $0.089\pm0.017$   & 0.083  \\
$n_s$              & & $0.969\pm0.014$   & 0.961  & & $0.969\pm0.014$   & 0.968  \\
$\ln[10^{10}A_s]$  & & $3.153\pm0.044$   & 3.165  & & $3.152\pm0.044$   & 3.154  \\
$\Omega_{\Lambda}$ & & $0.590-0.752$     & 0.733  & & $0.588-0.751$     & 0.719 \\
\cline{3-4} \cline{6-7}
$\Omega_{r_c}$     & & $<0.0178$         & 0.0001 & & $<0.0186$         & 0.0003 \\
$\Omega_m$         & & $0.248\pm0.019$   & 0.248  & & $0.247\pm0.018$   & 0.247  \\
$H_0$              & & $71.9\pm1.9$      & 72.3   & & $71.9\pm1.9$      & 72.5   \\
\cline{1-1} \cline{3-4} \cline{6-7}
$-2\Delta\ln L$ & & \multicolumn{2}{|c|}{0.20} & & \multicolumn{2}{|c|}{0.13} \\
\cline{1-1} \cline{3-4} \cline{6-7}
\end{tabular}
}
\caption{Same as Table~\ref{tab:res_flat_sdgp}, but for the flat sDGP+$\Lambda$ model, quoting one-sided 1D marginalized upper 95\% CL for $\Omega_{r_c}$ and 68\% MCI for $\Omega_{\Lambda}$.}
\label{tab:res_flat_sdgp+l}
\end{table}

\begin{table}[ht]
\resizebox{\hsize}{!}{
\centering
\begin{tabular}{|l|r@{}|c|c|r@{}|c|c|}
\hline
%\cline{1-1} \cline{3-4} \cline{6-7}
Parameters & & \multicolumn{2}{|c|}{nDGP} & & \multicolumn{2}{|c|}{nDGP (with gISW)} \\
\cline{1-1} \cline{3-4} \cline{6-7}
$100\Omega_b h^2$  & & $2.237\pm0.054$   & 2.245  & & $2.238\pm0.056$   & 2.254  \\
$\Omega_c h^2$     & & $0.1109\pm0.0049$ & 0.1095 & & $0.1100\pm0.0046$ & 0.1076 \\
$\theta$           & & $1.0406\pm0.0027$ & 1.0410 & & $1.0407\pm0.0027$ & 1.0409 \\
$\tau$             & & $0.084\pm0.016$   & 0.084  & & $0.085\pm0.017$   & 0.092  \\
$n_s$              & & $0.958\pm0.012$   & 0.961  & & $0.959\pm0.013$   & 0.961  \\
$\ln[10^{10}A_s]$  & & $3.196\pm0.043$   & 3.182  & & $3.191\pm0.043$   & 3.190  \\
$\Omega_{\Lambda}$ & & $0.754-0.934$     & 0.765  & & $0.753-0.924$     & 0.772 \\
\cline{3-4} \cline{6-7}
$\Omega_{r_c}$     & & $<0.0228$         & 0.0001 & & $<0.0203$         & 0.0001 \\
$\Omega_m$         & & $0.247\pm0.019$   & 0.253  & & $0.243\pm0.018$   & 0.244  \\
$H_0$              & & $73.6\pm2.0$      & 72.2   & & $73.9\pm2.0$      & 73.0   \\
\cline{1-1} \cline{3-4} \cline{6-7}
$-2\Delta\ln L$ & & \multicolumn{2}{|c|}{0.05} & & \multicolumn{2}{|c|}{0.23} \\
\cline{1-1} \cline{3-4} \cline{6-7}
\end{tabular}
}
\caption{Same as Table~\ref{tab:res_flat_sdgp+l}, but for the flat nDGP model.}
\label{tab:res_flat_ndgp}
\end{table}

We begin with the analysis of flat $\Lambda$CDM without DGP modifications
in Table~\ref{tab:res_flat_lcdm}.    We
show constraints with and without the gISW data and the maximum likelihood
parameters and value.
Horizontal lines divide the chain parameters from the derived parameters
and the best-fit (maximum) likelihood.
In the case of $\Lambda$CDM, the inclusion of the gISW data does not yield noticeable improvement on the parameter constraints~\cite{ho:08}.
This analysis sets the baseline
by which adding the DGP degrees of freedom should be measured.

In the flat sDGP model without $\Lambda$, there is no choice of parameters that can satisfy
the joint requirements of geometrical measurements from the CMB, supernovae, and $H_0$ and
the dynamical requirements from the ISW effect. For sDGP, we find $-2\Delta \ln L = 32.7$ with respect to $\Lambda$CDM and $-2\Delta \ln L = 33.1$ ($5.8\sigma$) when including the gISW likelihood.
In this case, the ISW effect is so large at low
multipoles
that the CMB alone rules out such contributions \cite{fang:08} and the gISW constraint adds
only an insignificant amount of extra information (see Tables~\ref{tab:res_flat_sdgp}).
The strengthening of the constraint when compared to 
Ref.~\cite{fang:08} comes from the improved Hubble constant measurements.

 In the sDGP+$\Lambda$ and nDGP models, the cosmological constant becomes a free parameter and we have to add it to the parameter set, hence, $P \rightarrow P \ \cup \ \{ \Omega_{\Lambda} \}$. $\Omega_{r_c} $ is a derived parameter and in particular we get $\Omega_{r_c}\rightarrow0$ in the limit $\Omega_{\Lambda}\rightarrow(1-\Omega_m)$.  In this limit, the phenomenology of
 $\Lambda$CDM is recovered for all observables.  Preference for a finite $\Omega_{r_c}$ indicates
 evidence for the DGP modification in these cases.

 In both the nDGP and sDGP+$\Lambda$ cases the maximum likelihood models differ insignificantly from $\Lambda$CDM (see Tables~\ref{tab:res_flat_sdgp+l} and \ref{tab:res_flat_ndgp}) and there
is no preference for finite $\Omega_{r_c}$.
Conversely, both branches require a finite $\Omega_\Lambda$ at high significance.

Since $\Lambda$CDM is the $\Omega_{r_c} \rightarrow 0$ limit of both branches with $\Lambda$,
the slightly poorer fit for nDGP and sDGP+$\Lambda$ should be attributed to sampling error in the MCMC.
The one-sided 1D marginalized upper 95\% confidence limits for $\Omega_{r_c}$ are
$\Omega_{r_c}<0.0178 (0.0186)$
for sDGP+$\Lambda$ and
 $\Omega_{r_c}< 0.0228 (0.0203)$ for nDGP where the values in parentheses include
 the gISW constraint.
 These values indicate that the crossover scale is at least substantially greater than the
 Hubble scale $H_0 r_c \simgt 3.5$.  
 
 In this $\Lambda$CDM limit, the modifications to
 the gISW predictions do not affect the constraints.
 The slight weakening 
of the constraints with the inclusion of gISW in sDGP+$\Lambda$ does not indicate a statistically significant
tension but does
suggest that future improvement in constraints can tighten the bounds on $H_0 r_c$. In particular, sDGP modifications tend to enhance correlations at high redshift
relative to low redshift.  The current data have a marginal preference for increased
correlation with redshift relative to $\Lambda$CDM (see Fig.~\ref{fig:ISW}).

 Note that due to the distinctive skewness of the posterior distribution, we give the 1D marginalized 68\% minimum credible intervals (MCI) (see Ref.~\cite{hamann:07}) for the brane tension $\Omega_{\Lambda}$ as opposed to the standard deviations given for the other parameters.

Finally, in the context of these flat models the possibility of phantom equations of state
currently is highly constrained.  For nDGP  $1+w_0 > -0.039$ at the 95\% C.L.

\subsection{Nonflat universe constraints}
\label{sec:nonflat}

In a universe with spatial curvature, we include $\Omega_K$ as a
parameter in the chain for each of the model classes. We use the
prior $\Omega_K\in(-0.1,0.1)$, which we weaken to
$\Omega_K\in(-1,1)$ in nDGP since we expect degeneracies between
$\Omega_K$ and $\Omega_{\Lambda}$. We also implement latter prior for
sDGP+$\Lambda$. For $\Lambda$CDM, Ho et al.~\cite{ho:08} have found
an improvement of the constraints on $\Omega_K$ by a factor of 3.2,
with respect to WMAP3 data {\it alone}, due to the inclusion of the
gISW and weak lensing data. However we find that the inclusion of
the other data, specifically the supernova and $H_0$ data, make
curvature constraints only marginally improved by the gISW
inclusion. We again use these $\Lambda$CDM results shown in
Table~\ref{tab:res_k_lcdm} as a baseline for comparison with sDGP,
sDGP+$\Lambda$, and nDGP in
 Tabsles~\ref{tab:res_k_sdgp}, \ref{tab:res_k_sdgp+l}, and \ref{tab:res_k_ndgp}.

For sDGP without $\Lambda$, adding curvature alleviates the tension between
CMB and supernova distance measures. However, it cannot reduce the ISW contributions
\cite{song:06,fang:08}
and so  we obtain $-2\Delta \ln L = 23.3 (23.8)$,  with respect to $\Lambda$CDM
where values in parentheses include the gISW constraint.   Utilizing all of the
data, the significance of the exclusion of sDGP without $\Lambda$ is
$\sim 5\sigma$.

\begin{figure*}
\resizebox{\hsize}{!}{\includegraphics{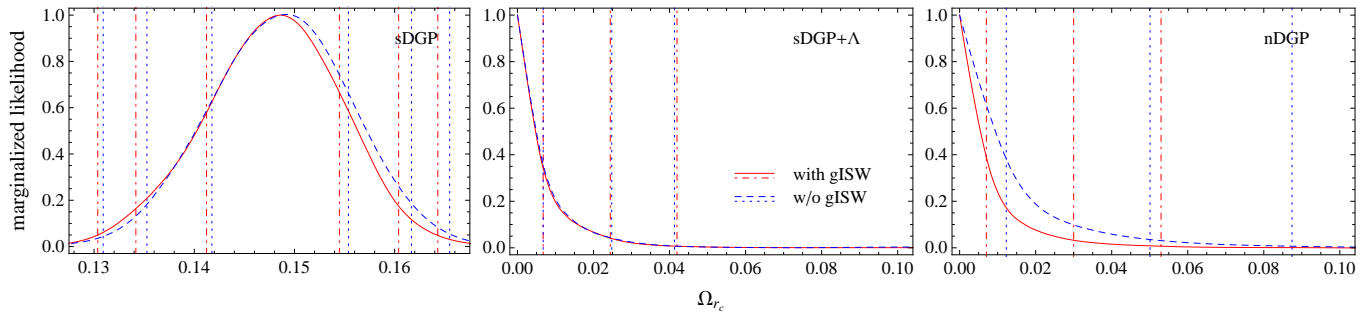}}
\caption{Marginalized likelihood for $\Omega_{r_c}$ in the nonflat nDGP and sDGP models. The vertical lines indicate 68\%, 95\%, and 99\% C.L.}
\label{fig:marglike}
\end{figure*}

Similar to the flat case, we find no preference for a finite $\Omega_{r_c}$ in nDGP and sDGP+$\Lambda$ and consequently no indications of DGP modifications to gravity (see Fig.~\ref{fig:marglike}).
With sDGP+$\Lambda$, we are again driven to the limiting case of $\Lambda$CDM with the slightly poorer best fit reflecting a sampling error in the chain.  Allowance for curvature on the
other hand weakens
the upper limit on the DGP modifications: $\Omega_{r_c}< 0.0248 (0.0244)$ and
$H_0 r_c > 3.18 (3.20)$ at 95\% C.L.

\begin{figure*}
\resizebox{\hsize}{!}{\includegraphics{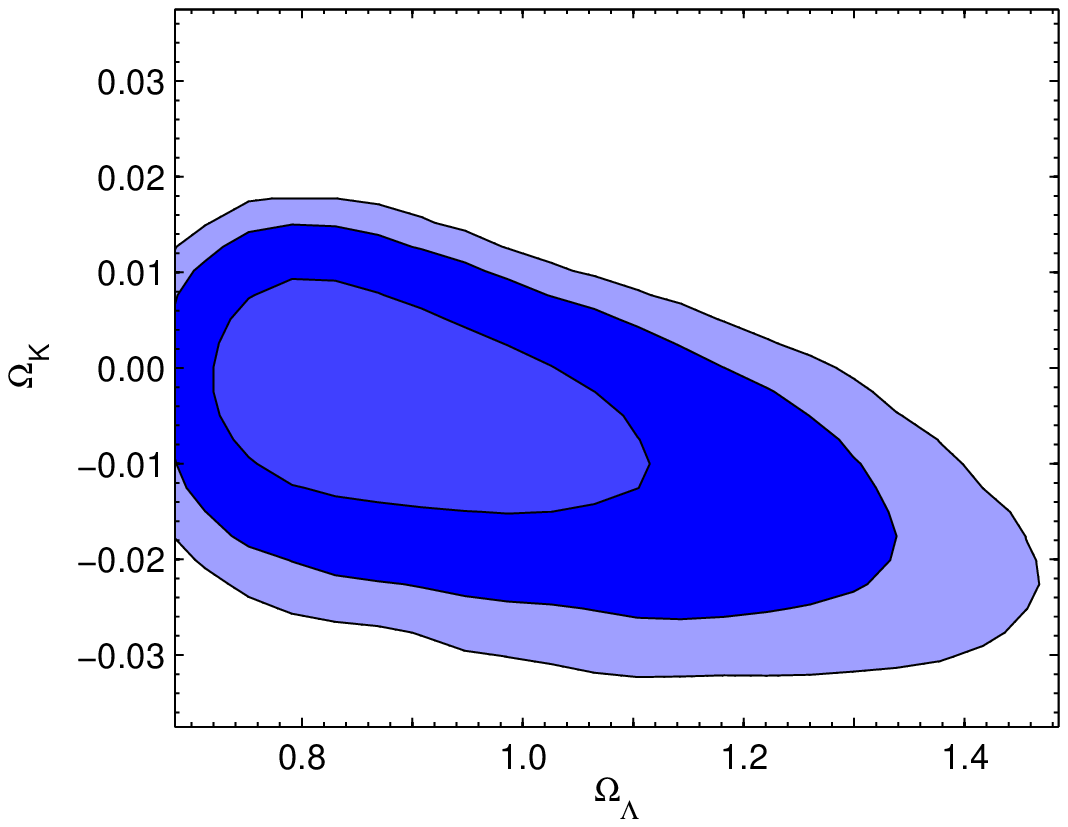} \ \ \includegraphics{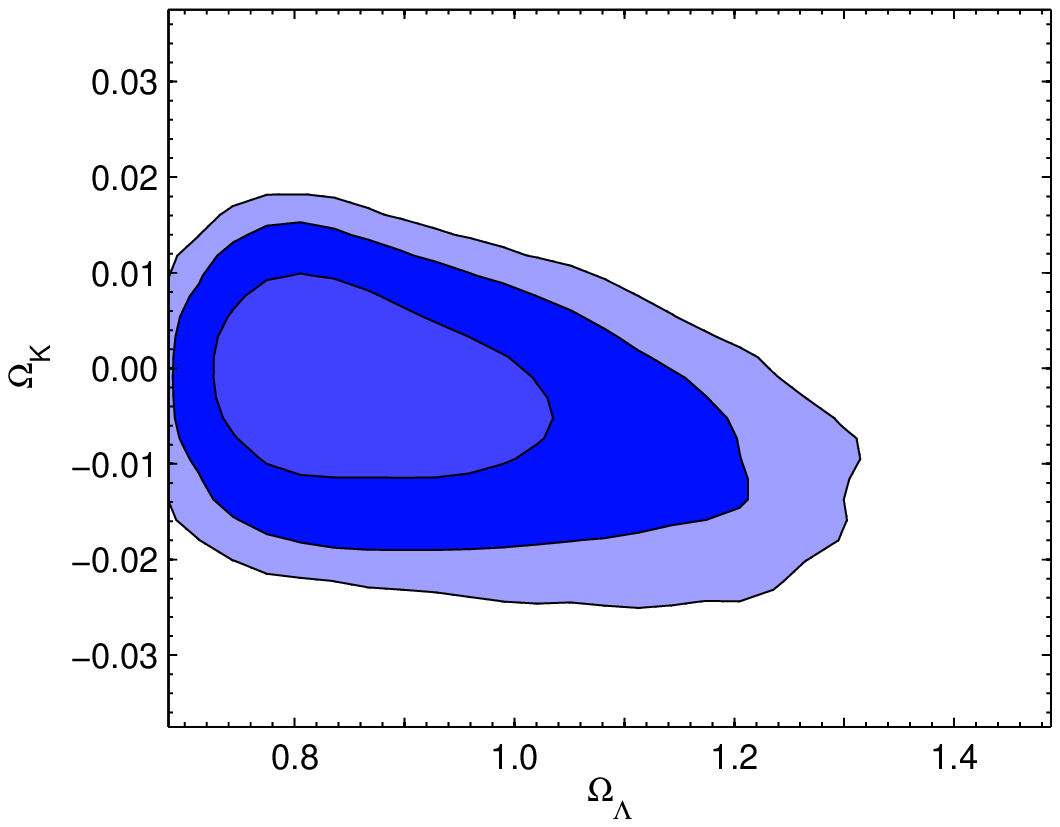}}
\caption{Contours of 2D marginalized 68\%, 95\%, and 99\% confidence boundaries using WMAP5, ACBAR, CBI, VSA, SNLS, and SHOES (left panel), including gISW (right panel) for nDGP.}
\label{fig:contours}
\end{figure*}

 For nDGP, the addition of curvature introduces a degeneracy with the cosmological constant.
 As was pointed out by Giannantonio et al.~\cite{giannantonio:08}, this degeneracy can be broken by the use of galaxy-ISW cross correlations since high curvature solutions underpredict
 the correlation especially at high redshift.
 Figure~\ref{fig:contours} illustrates this degeneracy and the effect of gISW measures.
The result of marginalizing curvature in nDGP is again a weakening of the DGP constraints 
 $\Omega_{r_c}< 0.0501 (0.0300)$ and
$H_0 r_c > 2.23 (2.89)$ at 95\% C.L.

In summary with the gISW constraint, the
limit on either branch implies $H_0 r_c \simgt 3$ and only a small weakening from the
flat case of $3.5$.
Furthermore  due to the curvature degeneracy in nDGP, restrictions on phantomlike equations of
state are also somewhat weakened to $w_0+1<-0.049$.

\begin{table}[ht]
\resizebox{\hsize}{!}{
\centering
\begin{tabular}{|l|r@{}|c|c|r@{}|c|c|}
\hline
%\cline{1-1} \cline{3-4} \cline{6-7}
Parameters & & \multicolumn{2}{|c|}{$\Lambda$CDM} & & \multicolumn{2}{|c|}{$\Lambda$CDM (with gISW)} \\
\cline{1-1} \cline{3-4} \cline{6-7}
$100\Omega_b h^2$  & & $2.250\pm0.056$    & 2.246   & & $2.249\pm0.055$    & 2.238   \\
$\Omega_c h^2$     & & $0.1084\pm0.0052$  & 0.1095  & & $0.1084\pm0.0051$  & 0.1085  \\
$\theta$           & & $1.0412\pm0.0027$  & 1.0412  & & $1.0411\pm0.0027$  & 1.0419  \\
$\tau$             & & $0.086\pm0.017$    & 0.090   & & $0.087\pm0.017$    & 0.083   \\
$n_s$              & & $0.963\pm0.013$    & 0.960   & & $0.963\pm0.013$    & 0.962   \\
$\ln[10^{10}A_s]$  & & $3.176\pm0.044$    & 3.196   & & $3.179\pm0.043$    & 3.174   \\
$\Omega_K$         & & $-0.0001\pm0.0063$ & 0.0020 & & $0.0007\pm0.0062$ & 0.0021 \\
\cline{3-4} \cline{6-7}
$\Omega_{\Lambda}$ & & $0.751\pm0.020$    & 0.751   & & $0.753\pm0.019$    & 0.758   \\
$\Omega_m$         & & $0.249\pm0.022$    & 0.246   & & $0.246\pm0.022$    & 0.240   \\
$H_0$              & & $72.6\pm3.0$       & 73.2    & & $73.0\pm3.0$       & 73.8    \\
\cline{1-1} \cline{3-4} \cline{6-7}
$-2\ln L$ & & \multicolumn{2}{|c|}{2834.01} & & \multicolumn{2}{|c|}{2867.74} \\
\cline{1-1} \cline{3-4} \cline{6-7}
\end{tabular}
}
\caption{$\Lambda$CDM as in Table~\ref{tab:res_flat_lcdm}, except allowing spatial curvature.}
\label{tab:res_k_lcdm}
\end{table}

\begin{table}[ht]
\resizebox{\hsize}{!}{
\centering
\begin{tabular}{|l|r@{}|c|c|r@{}|c|c|}
\hline
%\cline{1-1} \cline{3-4} \cline{6-7}
Parameters & & \multicolumn{2}{|c|}{sDGP} & & \multicolumn{2}{|c|}{sDGP (with gISW)} \\
\cline{1-1} \cline{3-4} \cline{6-7}
$100\Omega_b h^2$  & & $2.377\pm0.061$   & 2.365  & & $2.376\pm0.062$   & 2.352  \\
$\Omega_c h^2$     & & $0.0951\pm0.0041$ & 0.0970 & & $0.0952\pm0.0039$ & 0.0979 \\
$\theta$           & & $1.0441\pm0.0028$ & 1.0451 & & $1.0441\pm0.0028$ & 1.0439 \\
$\tau$             & & $0.091\pm0.020$   & 0.084  & & $0.092\pm0.019$   & 0.084  \\
$n_s$              & & $1.004\pm0.014$   & 1.002  & & $1.004\pm0.014$   & 0.997  \\
$\ln[10^{10}A_s]$  & & $3.018\pm0.043$   & 3.019  & & $3.021\pm0.043$   & 3.037  \\
$\Omega_K$         & & $0.0186\pm0.0055$ & 0.0212 & & $0.0182\pm0.0055$ & 0.0220 \\
\cline{3-4} \cline{6-7}
$\Omega_{r_c}$     & & $0.1486\pm0.0068$ & 0.1486 & & $0.1479\pm0.0067$ & 0.1467 \\
$\Omega_m$         & & $0.218\pm0.019$   & 0.216  & & $0.220\pm0.019$   & 0.220  \\
$H_0$              & & $74.0\pm3.0$      & 74.7   & & $73.7\pm2.9$      & 74.2   \\
\cline{1-1} \cline{3-4} \cline{6-7}
$-2\Delta\ln L$ & & \multicolumn{2}{|c|}{23.32} & & \multicolumn{2}{|c|}{23.79} \\
\cline{1-1} \cline{3-4} \cline{6-7}
\end{tabular}
}
\caption{sDGP without $\Lambda$ as in Table~\ref{tab:res_flat_sdgp}, except allowing spatial curvature.
$-2\Delta\ln L$ is quoted with respect to the maximum likelihood $\Lambda$CDM model with curvature here and in the following tables.}
\label{tab:res_k_sdgp}
\end{table}

\begin{table}[ht]
\resizebox{\hsize}{!}{
\centering
\begin{tabular}{|l|r@{}|c|c|r@{}|c|c|}
\hline
%\cline{1-1} \cline{3-4} \cline{6-7}
Parameters & & \multicolumn{2}{|c|}{sDGP+$\Lambda$} & & \multicolumn{2}{|c|}{sDGP+$\Lambda$ (with gISW)} \\
\cline{1-1} \cline{3-4} \cline{6-7}
$100\Omega_b h^2$  & & $2.266\pm0.058$    & 2.252  & & $2.266\pm0.0059$  & 2.251   \\
$\Omega_c h^2$     & & $0.1065\pm0.0051$  & 0.1066 & & $0.1064\pm0.0051$ & 0.1095  \\
$\theta$           & & $1.0416\pm0.0028$  & 1.0406 & & $1.0415\pm0.0028$ & 1.0414  \\
$\tau$             & & $0.087\pm0.017$    & 0.077  & & $0.088\pm0.017$   & 0.090   \\
$n_s$              & & $0.968\pm0.014$    & 0.962  & & $0.968\pm0.014$   & 0.960   \\
$\ln[10^{10}A_s]$  & & $3.157\pm0.046$    & 3.153  & & $3.158\pm0.046$   & 3.197   \\
$\Omega_K$         & & $0.0032\pm0.0068$ & 0.0022 & & $0.0036\pm0.0065$ & 0.0018 \\
$\Omega_{\Lambda}$ & & $0.557-0.745$      & 0.711  & & $0.561-0.746$ & 0.737   \\
\cline{3-4} \cline{6-7}
$\Omega_{r_c}$     & & $<0.0248$          & 0.0006 & & $<0.0244$         & 0.0000  \\
$\Omega_m$         & & $0.245\pm0.022$    & 0.240  & & $0.243\pm0.021$   & 0.248   \\
$H_0$              & & $72.8\pm3.0$       & 73.3   & & $73.1\pm2.9$      & 72.9    \\
\cline{1-1} \cline{3-4} \cline{6-7}
$-2\Delta\ln L$ & & \multicolumn{2}{|c|}{0.07} & & \multicolumn{2}{|c|}{0.04} \\
\cline{1-1} \cline{3-4} \cline{6-7}
\end{tabular}
}
\caption{sDGP with $\Lambda$ as in Table~\ref{tab:res_flat_sdgp+l}, but allowing spatial curvature.}
\label{tab:res_k_sdgp+l}
\end{table}

\begin{table}[ht]
\resizebox{\hsize}{!}{
\centering
\begin{tabular}{|l|r@{}|c|c|r@{}|c|c|}
\hline
%\cline{1-1} \cline{3-4} \cline{6-7}
Parameters & & \multicolumn{2}{|c|}{nDGP} & & \multicolumn{2}{|c|}{nDGP (with gISW)} \\
\cline{1-1} \cline{3-4} \cline{6-7}
$100\Omega_b h^2$  & & $2.239\pm0.056$   & 2.245  & & $2.242\pm0.056$    & 2.239   \\
$\Omega_c h^2$     & & $0.1099\pm0.0054$ & 0.1076 & & $0.1094\pm0.0054$  & 0.1099  \\
$\theta$           & & $1.0409\pm0.0027$ & 1.0412 & & $1.0409\pm0.0027$  & 1.0409  \\
$\tau$             & & $0.084\pm0.017$   & 0.084  & & $0.085\pm0.017$    & 0.091   \\
$n_s$              & & $0.959\pm0.013$   & 0.960  & & $0.960\pm0.013$    & 0.960   \\
$\ln[10^{10}A_s]$  & & $3.189\pm0.045$   & 3.176  & & $3.188\pm0.045$    & 3.205   \\
$\Omega_K$         & & $-0.0055\pm0.0080$  & -0.0056 & & $-0.0029\pm0.0069$ & 0.0021 \\
$\Omega_{\Lambda}$ & & $0.749-1.009$     & 0.801  & & $0.749-0.953$      & 0.764    \\
\cline{3-4} \cline{6-7}
$\Omega_{r_c}$     & & $<0.0501$         & 0.0008  & & $<0.0300$          & 0.0000  \\
$\Omega_m$         & & $0.255\pm0.023$   & 0.261  & & $0.248\pm0.022$    & 0.247   \\
$H_0$              & & $72.1\pm3.0$      & 70.6   & & $73.0\pm3.0$       & 73.1    \\
\cline{1-1} \cline{3-4} \cline{6-7}
$-2\Delta\ln L$ & & \multicolumn{2}{|c|}{0.09} & & \multicolumn{2}{|c|}{0.41} \\
\cline{1-1} \cline{3-4} \cline{6-7}
\end{tabular}
}
\caption{nDGP as in Table~\ref{tab:res_flat_ndgp}, but allowing spatial curvature.}
\label{tab:res_k_ndgp}
\end{table}

\section{Discussion}\label{sec:discussion}

We have performed the first Markov chain Monte Carlo analysis of the nDGP and sDGP branches of DGP braneworld gravity to utilize all of the CMB data, including
the lowest multipoles, and its correlation
with galaxies (gISW).   We also include supernovae and Hubble constant data in the
constraint.

We find no preference for DGP modifications to gravity on either branch.
Indeed, on the self-accelerating branch without $\Lambda$, the model is excluded
at the $4.9\sigma$ and $5.8\sigma$ levels with and without curvature respectively
\cite{fang:08}.
While the gISW data do not substantially improve this constraint, they do additionally
disfavor sDGP.

With the inclusion of $\Lambda$ on either branch, the DGP model cannot be entirely
excluded but its modifications are strongly limited.  We find that the crossover
scale, which measures the strength of the modifications, must be substantially above
the Hubble scale $H_0 r_c > 3$ with curvature and $3.5$ without curvature.
The robustness of this constraint is substantially assisted by the gISW data.
In nDGP, it breaks the geometric degeneracy between $\Lambda$ and spatial curvature.
In sDGP, the relatively large correlation at high redshift offers opportunities in the 
future for improving the limits on $H_0 r_c$.
These abilities highlight the importance of obtaining improved gISW data for constraining
infrared modifications to gravity.

\section*{Acknowledgments}

We would like to thank Kazuya Koyama, Sanjeev Seehra, Fabian
Schmidt, and Yong-Seon Song for useful discussions and An\v{z}e Slosar for helpful insights into the CosmoMC and ISWWLL codes. Computational resources were provided on the zBox2 supercomputer at the University of Z\"{u}rich. This work was partially supported by the
Swiss National Foundation
under Contract No.~200021-116696/1 and WCU Grant No.~R32-2008-000-10130-0.
W.H. was supported by the Kavli Institute for Cosmological
 Physics (KICP) at the University of Chicago through Grants NSF No.~PHY-0114422 and
 NSF No.~PHY-0551142, U.S. Department of Energy Contract No.~DE-FG02-90ER-40560, and
 the David and Lucile Packard Foundation. W.F. was supported by the U.S. Department of Energy Contract No.~DE-AC02-98CH10886.
\appendix

\section{Modifications to the ISWWLL code}

We use the publicly available ISWWLL code~\cite{ho:08, hirata:08}
for our analysis. Note that we have turned off weak lensing likelihood contributions in the code, focusing only on the gISW constraints. The 42 data points of gISW cross correlations that are used in the likelihood analysis are collected from the
Two Micron All Sky Survey (2MASS) extended source catalog (XSC)~\cite{jarrett:00, skrutskie:06}, the luminous red galaxies (LRG) and photometric quasars (QSO) of the Sloan Digital Sky Survey (SDSS)~\cite{adelman:07}, and the National Radio Astronomy Observatory (NRAO) Very Large Array (VLA) Sky Survey (NVSS)~\cite{condon:98}. They are divided into nine galaxy sample bins $j$ (2MASS0-3, LRG0-1, QSO0-1, NVSS) based on flux (2MASS) or redshift (LRG, QSO). These data points are a selection of multipole bins from all samples, where the selection is based on the avoidance of nonlinearities and systematic effects from dust extinction, galaxy foregrounds, the thermal Sunyaev-Zel'dovich effect, and point source contamination to affect the gISW cross correlations~\cite{ho:08}.

In the remainder of this Appendix, we discuss the details of the modifications implemented in the ISWWLL code. 
First,
we describe the calculation of
 the quasistatic linear growth rate  $D(z)$ in the gISW cross correlation, Eq.~(\ref{eq:ClgT}), and analyze the validity of the Limber and the quasistatic approximation.
 We then discuss the function $f_j(z)$ that carries information about the redshift distribution and bias.

\subsection{gISW cross correlations}

It has been argued that for nDGP and sDGP the gISW cross correlations are well described within the quasistatic regime~\cite{song:06, giannantonio:08, koyama:05, sawicki:06}. Here, this can easily be seen from the substitution $k\rightarrow(\ell+1/2)/\chi(z)$ considering the relevant redshifts. In this limit, we solve the ordinary differential equation \cite{lue:04}
\begin{equation}
\Delta_m'' + \left( 2+\frac{H'}{H} \right) \Delta_m' - \frac{3}{2} \left( 1-g_{\rm QS} \right) \frac{H_0^2\Omega_m}{a^3H^2} \Delta_m = 0
\label{eq:Delta_ODE}
\end{equation}
for the linear matter density perturbation $\Delta_m$. Note that for nDGP, in the limit $r_c \rightarrow \infty$, we have $g_{\rm QS} \rightarrow 0$ and $H(z)$ approaches the expansion history of $\Lambda$CDM. Therefore, in this limit, Eq.~(\ref{eq:Delta_ODE}) recovers the quasistatic ordinary differential equation for the matter overdensity in $\Lambda$CDM. We solve Eq.~(\ref{eq:Delta_ODE}) with initial conditions at $a_i \ll 1$, in a regime where general relativity is expected to hold, i.e., $\Delta_m'(a_i)=\Delta_m(a_i)$ with a normalization set by the
initial power spectrum.

The accuracy of the Limber approximation in the case of $\Lambda$CDM is at the order of 10\% at $\ell=2$ and drops approximately as $\ell^2$ at higher $\ell$ (see e.g.~\cite{smith:09, loverde:08, afshordi:04}). The error depends further on the width of the redshift distribution, which changes only little with DGP effects. The relative deviation from the exact result at $\ell=6$ does not exceed $\sim3\%$ for the samples and typical models considered in Fig.~\ref{fig:ISW}. Given the large errors of the currently available data points at low $\ell$, we conclude that the Limber approximation is applicable and furthermore very useful since it is numerically faster than an exact integration.

\subsection{Redshift distribution and bias}

A further modification to the code that we need to conduct is in the determination of the function $f_j(z)$. In the Markov chain, $f_j(z)$ is recomputed when changing the cosmological parameter values. The methods by which this function is determined differ for each sample, but they are all based on galaxy clustering data.

The 2MASS galaxies are matched with SDSS galaxies in order to identify their redshifts. To obtain the nonlinear power spectrum, the $Q$ model for nonlinearities~\cite{cole:05} is applied. Then, the code computes the galaxy power spectrum and fits it to measurements, thereby determining the bias $b(z)$ and $Q$. Since the required accuracy for the estimation of bias is only at the order of a few tens of percent~\cite{ho:08}, this processing is also applicable to DGP. The $Q$ model is also adopted for LRG galaxies, where the redshift probability distribution is inferred with methods described in Ref.~\cite{padmanabhan:05}. For QSO, first, a preliminary estimate for the redshift distribution is deduced by locating a region of sky with high spectroscopic completeness, but simultaneously maintaining a large area. Taking into account magnification bias and fitting $b_j(z)\Pi_j(z)$ using the quasar power spectrum and quasar-LRG cross power yields the desired shape of $f_j(z)$. Finally, the effective redshift distribution of NVSS is obtained from cross-correlating with the other samples and $f_j(z)$ is fitted with a $\Gamma$ distribution.

The part of the ISWWLL code that is devoted to this processing is configured for a parametrization of the expansion history by $w_e = w_0 + (1-a)w_a$. This approach gives a good
approximation to sDGP in the domain of interest, but it fails for nDGP and sDGP+$\Lambda$ due to the appearance of a divergence in $w_e(a)$.  Therefore, instead of taking $w_0$ and $w_a$ to describe the expansion history, we utilize $\Omega_{r_c}$ and $\Omega_{\Lambda}$, where only the latter really is a necessary, free parameter.

In case of the SDSS quasars, the derivation of $f_j(z)$ involves the linear growth factor, which we need to replace by its DGP counterpart. This implies solving Eq.~(\ref{eq:Delta_ODE}). A further contribution for the QSO samples is due to magnification bias.  In the quasistatic regime of DGP
the relationship between the metric combination sensitive to gravitational redshifts and lensing ($\Phi-\Psi$)
and the density perturbations is unmodified so the expression of the lensing window function
for magnification effects given in Ref.~\cite{ho:08} is unchanged.
\vfill

%%%%%%%%%%%%%%%%%%%%%%%%%%%%%%%%%%%%%%%%%
\bibliographystyle{arxiv_physrev}

\bibliography{ndgp}
%%%%%%%%%%%%%%%%%%%%%%%%%%%%%%%%%%%%%%%%%

\end{document}